\documentstyle[12pt]{article}

\let\ni=\noindent

\begin{document}
\hfill IFT/99-21

\bigskip
\baselineskip 0.75cm

\renewcommand{\thefootnote}{\fnsymbol{footnote}}

\newcommand{\CKM}{Cabibbo--Kobayashi--Maskawa }

\newcommand{\SM}{Standard Model }

\pagestyle {plain}

\setcounter{page}{1}

\pagestyle{empty}

~~~

\vspace{0.3cm}

{\large\centerline{\bf An alternative for righthanded neutrinos:}}

{\large\centerline{\bf lefthanded see--saw{\footnote{Supported in part by the 
Polish KBN--Grant 2 P03B 052 16 (1999--2000).}}}}

\vspace{0.8cm}

{\centerline {\sc Wojciech Kr\'{o}likowski}}

\vspace{0.8cm}

{\centerline {\it Institute of Theoretical Physics, Warsaw University}}

{\centerline {\it Ho\.{z}a 69,~~PL--00--681 Warszawa, ~Poland}}

\vspace{0.5cm}

{\centerline{\bf Abstract}}

\vspace{0.3cm}

A new lefthanded see--saw mechanism is constructed, implying both the smallness
of active--neutrino masses and decoupling of heavy passive neutrinos, similarly
to the situation in the case of conventional see--saw. But now, in place of 
the conventional righthanded neutrinos, the lefthanded sterile neutrinos play 
the role of heavy passive neutrinos, the righthanded neutrinos and righthanded 
sterile neutrinos being absent. In this case, the neutrino mass term is 
necessarily of pure Majorana type.

\vspace{0.2cm}

\ni PACS numbers: 12.15.Ff , 14.60.Pq , 12.15.Hh .

\vspace{0.5cm}

\ni September 1999

\vfill\eject

~~~
\pagestyle {plain}

\setcounter{page}{1}

 As is well known, the popular see--saw mechanism [1] provides us with a gentle
way of introducing small neutrino masses into the original \SM, where Dirac--%
type masses are zero due to the absence of righthanded neutrinos. In fact, the 
actual righthanded neutrinos (of all three generations) get in this mechanism 
large (by assumption) Majorana--type masses and so, become practically 
decoupled from lefthanded neutrinos that are allowed to carry only small 
Majorana--type masses. Such large  righthanded Majorana masses are, on the 
other hand, related to an expected high energy scale at which lepton number
violation ought to appear.

 In this note, we present a new, {\it a priori} possible mechan\-ism, where the
role played in the see--saw mechanism by right\-handed neutrinos is taken over 
by hypothetic {\it lefthanded} sterile neutrinos free of any \SM charges. Thus,
the new mechanism may be called {\it lefthanded see--saw}. In this case, both 
righthanded neutrinos and righthanded sterile neutrinos are conjectured to be 
absent, $\nu_R \equiv 0 $ and $\nu^{(s)}_R \equiv 0 $, so that

\vspace{-0.1cm}

%rownanie 1
\begin{equation}
\nu \equiv \nu_L \;\;,\;\;\nu^{(s)} \equiv \nu^{(s)}_L
\end{equation}

\vspace{-0.1cm}

\ni are active and sterile neutrinos, respectively (for all three neutrino 
generations).

 In the new situation, the neutrino mass term is of pure Majorana type

\vspace{-0.1cm}

%rownanie 2
\begin{eqnarray}
-{\cal L}_{\rm mass} & = & \frac{1}{2}\left(\overline{(\nu_L)^c}\,,\,\overline{
(\nu_L^{(s)})^c} \right) \left(\begin{array}{cc} m^{(L)} & \mu^{(L)} 
\\ \mu^{(L)}_s & m^{(L)}_s \end{array} \right) \left(\begin{array}{c} \nu^{L} 
\\ \nu^{(s)}_L \end{array} \right) + {\rm h.c.} \nonumber \\ & = & \frac{1}{2}
\left(\overline{(\nu_M)^c}\,,\,\overline{(\nu_M^{(s)})^c} \right) \left(
\begin{array}{cc} m^{(L)} & \mu^{(L)} \\ \mu^{(L)}_s & m^{(L)}_s \end{array} 
\right) \left(\begin{array}{c} \nu^{M} \\ \nu^{(s)}_M \end{array} \right) \;,
\end{eqnarray}

\vspace{-0.1cm}

\ni where

\vspace{-0.1cm}

%rownanie 3
\begin{equation}
\nu_M \equiv \nu_L + (\nu_L)^c\;\;,\;\; \nu^{(s)}_M \equiv \nu^{(s)}_L +
(\nu^{(s)}_L)^c
\end{equation}

\ni (in the case of one neutrino generation). In the case of three neutrino 
generations, real numbers $ m^{(L)} $, $ m^{(L)}_s $ and $ \mu^{(L)} $ 
become $ 3\times 3 $ matrices $ \widehat{m}^{(L)}_s $, $ \widehat{m}^{(L)}_s $
and $ \widehat{\mu}^{(L)}$, real and symmetric for simplicity, while neutrino 
fields $ \nu_L $ and $ \nu^{(s)}_L $ transit into the field columns $ \vec{\nu
}_L = \left(\nu_{e\,L}\,,\,\nu_{\mu\,L}\,,\,\nu_{\tau\,L} \right)^T $ and $ 
\vec{\nu}^{(s)}_L = \left(\nu^{(s)}_{e\,L}\,,\,\nu^{(s)}_{\mu\,L}\,,\,\nu^{(s)
}_{\tau\,L} \right)^T $.

 In Eq. (2), all four terms violate lepton number ($\Delta L = \pm 2 $), the 
terms proportional to $\mu^{(L)}$ and $ m^{(L)}$ break electroweak symmetry, 
while the term proportional to $m_s^{(L)}$ is electroweak gauge invariant. The 
terms $\mu^{(L)}$ and $ m^{(L)}$ may be generated spontaneously by Higgs 
mechanism, the first --- by the linear Higgs coupling (of the Majorana type):

\vspace{-0.1cm}

%rownanie 4
\begin{equation}
{\cal L}_H = \frac{1}{2} g_H \left[ \overline{(l_L)^c} H \nu^{(s)}_L - 
\overline{l_L} H^c (\nu^{(s)}_L)^c \right] + 
{\rm h.c.}\;\;,\;\;\mu^{(L)} \equiv g_H \langle H^\circ \rangle
\end{equation}

\ni and the second --- by the familiar bilinear effective Higgs coupling (also 
of the Majorana type):

\vspace{-0.1cm}

%rownanie 5
\begin{equation}
{\cal L}_{HH} = \frac{1}{4 M} g_{HH} \left[\overline{(l_L)^c}\vec{\tau} l_L 
\right] \cdot \left( H^T i \tau_2 \vec{\tau} H\right) + {\rm h.c.}\;\;,\;\;
m^{(L)} \equiv \frac{1}{M} g_{HH} \langle H^\circ \rangle^2
\end{equation}

\ni (see {\it e.g.} Ref. [2]). Here, $\overline{(l_L)^c} = (l_L)^T C^{-1} i 
\tau_2 $ and

\vspace{-0.1cm}

%rownanie 6
\begin{eqnarray}
l_L = \left(\begin{array}{c} \nu_L \\ l^-_L \end{array} \right) & , & H = 
\left(\begin{array}{c} H^+ \\ H^\circ \end{array}\right)\;, \nonumber \\ 
(l_L)^c = i \tau_2 \left( \begin{array}{c} (\nu_L)^c \\ (l^-_L)^c \end{array}
\right) = \left( \begin{array}{c} (l^-_L)^c \\ - (\nu_L)^c \end{array} \right)
& , & H^c = i \tau_2 \left( \begin{array}{c} H^{+\,c} \\ H^{\circ\,c} 
\end{array}\right) = \left( \begin{array}{c} H^{\circ\,c} \\ -H^{+\,c}
\end{array}\right)\;,
\end{eqnarray}

\ni with $(\nu_L)^c = C\overline{\nu}_L^T $, $(l^-_L)^c = C\overline{l^{-\,T}_L
}$ and $H^{+\,c} = H^{+\,\dagger} = H^-$, $H^{\circ\,c} = H^{\circ\,\dagger}$ 
($H^{c\,\dagger} = - H^{T} i \tau_2 $). In Eq. (5), $ M $ is a large mass
scale probably related to the GUT scale, so that the inequality $ m^{(L)} \ll 
\mu^{(L)}$ is plausible. Note that the lefthanded sterile neutrino $\nu_L^{(s)}
$, a \SM scalar, may be an SU(5) scalar. However, it cannot be an SU(10) covar%
iant (say, a scalar or a member of 16--plet), since the SU(10) formula $ Y = 2 
I^{(R)}_3 + B - L $ for weak hypercharge does not work in the case of $\nu_L^{
(s)}$ with $ L = 1 $ and $ B = 0 $ ($ Y = 0 $ and $ I^{(R)}_3 = 0 $ imply 
$ B - L = 0 $). Thus, the existence of $\nu_L^{(s)}$ breaks dynamically the 
SU(10) symmetry, unless $\nu_L^{(s)}$ gets $ B - L = 0 $ ({\it e.g.} $ L = 0 
= B $ or $ L = 1 = B $).

 After its diagonalization, the mass term (2) becomes

\vspace{-0.1cm}

%rownanie 7
\begin{equation}
-{\cal L}_{\rm mass} = \frac{1}{2}\left(\overline{\nu}_I\,,\,\overline{
\nu}_{II} \right) \left(\begin{array}{cc} m_I & 0 \\ 0 & m_{II} \end{array} 
\right) \left( \begin{array}{c} \nu_I \\ \nu_{II} \end{array} \right) \;,
\end{equation}

\ni where 

\vspace{-0.1cm}

%rownanie 8
\begin{eqnarray}
\nu_{I} & = & \nu_M \cos \theta - \nu^{(s)}_M \sin \theta\;, \nonumber \\ 
\nu_{I} & = & \nu_M \sin \theta + \nu^{(s)}_M \cos \theta
\end{eqnarray}

\ni with $\tan \theta = (m_{II} - m^{(L)}_s)/m^{(L)}_s $, and

\vspace{-0.1cm}

%rownanie 9
\begin{equation}
m_{I,\,II} = \frac{m^{(L)} + m^{(L)}_s}{2} \mp \sqrt{\left( \frac{m^{(L)} - 
m^{(L)}_s}{2}\right)^2 + \mu^{(L)\,2}}\;.
\end{equation}

\ni Assuming that

\vspace{-0.1cm}

%rownanie 10
\begin{equation}
0 \leq m^{(L)} \ll \mu^{(L)} \ll m^{(L)}_s\;,
\end{equation}

\vspace{-0.1cm}

\ni we obtain the neutrino mass eigenstates 

\vspace{-0.1cm}

%rownanie 11
\begin{equation}
\nu_I \simeq \nu_M - \frac{\mu^{(L)}}{m^{(L)}_s} \nu^{(s)}_M \simeq \nu_M \;,\;
\nu_{II} \simeq \nu^{(s)}_M - \frac{\mu^{(L)}}{m^{(L)}_s} \nu_M \simeq 
\nu_M^{(s)} 
\end{equation}

\vspace{-0.1cm}

\ni related to the neutrino masses

\vspace{-0.1cm}

%rownanie 12
\begin{equation}
m_I \simeq  - \frac{\mu^{(L)\,2}}{m^{(L)}_s}\;,\;m_{II} \simeq m^{(L)}_s\;.
\end{equation}

\vspace{-0.1cm}

\ni Thus, $|m_I| \ll m_{II}$, what practically decouples $\nu_L^{(s)}$ 
from $\nu_L $. Here, the minus sign at $ m_I $ is evidently irrevelant for $
\nu_I $ which, as a relativistic particle, is kinematically characterized by $
m^2_I $.

 New Eqs. (10) --- (12) give us a purely lefthanded counterpart of the popular 
see--saw mechanism, where the assumption 

\vspace{-0.1cm}

%rownanie 13
\begin{equation}
0 \leq m^{(L)} \ll m^{(D)} \ll m^{(R)}
\end{equation}

\vspace{-0.1cm}

\ni implies the neutrino mass eigenstates

\vspace{-0.2cm}

%rownanie 14
\begin{equation}
\nu_I \simeq \nu_M \equiv \nu_L +(\nu_L)^c \;,\;\nu_{II} \simeq \nu'_M \equiv
\nu_R + (\nu_R)^c
\end{equation}

\vspace{-0.1cm}

\ni connected with the neutrino masses 

\vspace{-0.1cm}

%rownanie 15
\begin{equation}
m_I \simeq  - \frac{m^{(D)\,2}}{m^{(R)}} \;,\; m_{II} \simeq 
m^{(L)}_s\;.
\end{equation}

\ni In that case, the neutrino mass term has the form

%rownanie 16
\begin{eqnarray}
-{\cal L}^{\rm conv}_{\rm mass} & = & \frac{1}{2}\left( \overline{(\nu_L)^c}
\,,\,\overline{\nu_R} \right) \left(\begin{array}{cc} m^{(L)} & m^{(D)} \\ m^{
(D)} & m^{(R)} \end{array} \right) \left( \begin{array}{c} \nu_L \\ (\nu_R)^c 
\end{array} \right) + {\rm h.c.} \nonumber \\ & = & \frac{1}{2}
\left( \overline{\nu_M}\,,\,\overline{\nu'_M} \right) \left( \begin{array}{cc} 
m^{(L)} & m^{(D)} \\ m^{(D)} & m^{(R)} \end{array} \right) \left(\begin{array}{
c} \nu_M \\ \nu'_M \end{array} \right) \;.
\end{eqnarray}

\vspace{-0.1cm}

\ni So, from Eq. (15) $|m_I| \ll m_{II}$, what leads to practically decoupling 
$\nu_R $ from $\nu_L $. But, while in Eq. (15) the magnitude of neutrino Dirac 
mass $ m^{(D)}$ may be compared with the mass of corresponding charged lepton, 
in Eq. (12) the magnitude of $\mu^{(L)}$, responsible for the coupling $(1/2)
\mu^{(L)}\left[\overline{(\nu_L)^c} \nu_L^{(s)} + \overline{\nu_L}(\nu_L^{(s)}
)^c \right] $ + h.c., may be quite different (perhaps smaller).

 In the general case of three neutrino generations, Eqs. (11) and (12) are 
replaced by

%\vspace{-0.1cm}

%rownanie 17
\begin{equation} 
\vec{\nu}_I \simeq \vec{\nu}_M \equiv \vec{\nu}_L + (\vec{\nu}_L)^c \;,\;
\vec{\nu}_{II} \simeq \vec{\nu}^{(s)}_M \equiv \vec{\nu}_L^{(s)} + (\vec{\nu
}_L^{(s)})^c
\end{equation}

\ni and

%rownanie 18
\begin{equation}
\widehat{m}_I \simeq  - \widehat{\mu}^{(L)}\left(\widehat{m}^{(L)}_s \right)^{
-1} \widehat{\mu}^{(L)}\;,\;\widehat{m}_{II} \simeq \widehat{m}^{(L)}_s
\end{equation}

\ni (compare {\it e.g.} Ref. [3]). Here, $\vec{\nu}_L = (\nu_{\alpha\,L})$,
$\vec{\nu}_I = (\nu_{I\,\alpha})$, $\widehat{m}_I = (m_{I\,\alpha \beta})$, 
{\it etc.} ($\alpha\,,\,\beta = e\,,\,\mu\,,\,\tau $). The Hermitian 
mass matrices $\widehat{m}_I $ and $\widehat{m}_{II}$ ($\widehat{m}^{(L)}$,
$\widehat{m}^{(L)}_s $ and $\widehat{\mu}^{(L)}$ were taken real and symmetric
for simplicity) can be diagonalized with the use of unitary matrices 
$\widehat{U}_I = \left( U_{I\,\alpha\,i}\right)$ and $\widehat{U}_{II} = \left(
U_{II\,\alpha\,i}\right)$, respectively, giving the neutrino mass eigenstates

%\vspace{-0.1cm}

%rownanie 19
\begin{eqnarray}
\nu_{I\,i} & = & \sum_\alpha\left(\widehat{U}_{I}^\dagger \right)_{i\,\alpha}
\nu_{I\,\alpha}  = \sum_\alpha U^*_{I\,\alpha\,i} \nu_{I\,\alpha}\;, \nonumber
\\ \nu_{II\,i} & = & \sum_\alpha\left(\widehat{U}_{II}^\dagger \right)_{i\,
\alpha} \nu_{II\,\alpha}  = \sum_\alpha U^*_{II\,\alpha\,i} \nu_{II\,\alpha}
\end{eqnarray}

\ni and the corresponding neutrino masses

%rownanie 20
\begin{eqnarray}
\delta_{ij} m_{I\,i} & = & \sum_{\alpha \beta} \left(\widehat{U}_{I}^\dagger 
\right)_{i\,\alpha} m_{I\,\alpha \beta}\left(\widehat{U}_{I} \right)_{\beta\,
j} = \sum_{\alpha \beta} U^*_{I\,\alpha\,i} U_{I\,\beta\,j} m_{I\,\alpha
\beta}\;, \nonumber \\ \delta_{ij} m_{II\,i} & = & \sum_{\alpha \beta} \left(
\widehat{U}_{II}^\dagger \right)_{i\,\alpha} m_{II\,\alpha \beta} \left( 
\widehat{U}_{II} \right)_{\beta\,j} = \sum_{\alpha \beta} U^*_{II\,\alpha\,i} 
U_{II\,\beta\,j} m_{II\,\alpha \beta}
\end{eqnarray}

\ni ($i,j = 1,2,3$). In our case, $ U_{I \alpha i}$  and $ U_{II \alpha i}$ 
are real, while $ m_{I\,\alpha \beta} \simeq -\sum_{\gamma \delta} 
\mu^{(L)}_{\alpha \gamma}$ $(\widehat{m}^{(L)\,-1}_s)_{\gamma \delta}\mu^{(L)
}_{\delta \beta}$ and $ m_{II\,\alpha \beta} \simeq m^{(L)}_{s \alpha \beta}$ 
due to Eqs. (18). When deriving Eq. (19), we assume that in the original 
lepton Lagrangian the charged--lepton mass matrix is diagonal and so, its 
diagonalizing unitary matrix $\widehat{U}^{(l)}$ is trivially equal to the unit
matrix. Then, the lepton counterpart of the \CKM matrix $\widehat{V} =\widehat{
U}^{(\nu)\,\dagger}\widehat{U}^{(l)}$ becomes equal to $\widehat{U}^{(\nu)\,
\dagger} = \widehat{U}^{\dagger}_I $, thus $V_{i \alpha} = U^*_{I \alpha i}$.

 From Eqs. (17) and (19) we conclude for neutrino fields that 

%rownanie 21
\begin{eqnarray}
\nu_{\alpha\,L} & \simeq & \left( \nu_{I\,\alpha} \right)_L = \sum_i 
U_{I\,\alpha\,i} \left(\nu_{I\,i} \right)_L \;, \nonumber \\
\nu_{\alpha\,L}^{(s)} & \simeq & \left( \nu_{II\,\alpha} \right)_L = \sum_i 
U_{II\,\alpha\,i} \left(\nu_{II\,i} \right)_L \;.
\end{eqnarray}

\ni Thus, the oscillation probabilities (on the energy shell) for active--%
neutrino states read (in the vacuum):

%rownanie 22
\begin{equation}
P(\nu_\alpha \rightarrow \nu_\beta) = |\langle \nu_\beta| e^{iPL}|
\nu_\alpha\rangle|^2 \simeq \delta_{\alpha \beta} - 4 \sum_{i<j} U^*_{I \beta 
j} U_{I \alpha j}U_{I \beta i} U^*_{I \alpha i} \sin^2 \left(1.27 \frac{m^2_{
I\,j} - m^2_{I\,i}}{E} L \right)\;,
\end{equation}

\ni where $ E = \sqrt{p^2_i + m^2_{I\,i}}$ ($i = 1,2,3$).

 If the matrix $\widehat{m}^{(L)}_s $ happens to be nearly diagonal and has 
nearly degenerate eigenvalues: $ m^{(L)}_{s \alpha \beta} \simeq \delta_{\alpha
\beta} m^{(L)}_s $ , then

%\vspace{-0.2cm}

%rownanie 23
\begin{equation}
m_{I \alpha \beta} \simeq - \sum_\gamma \frac{\mu^{(L)}_{\alpha \gamma}\mu^{(L)
}_{\gamma \beta}}{m^{(L)}_s}\;,\; m_{II \alpha \beta} \simeq \delta_{\alpha 
\beta} m^{(L)}_s 
\end{equation}

%\vspace{-0.1cm}

\ni and Eqs. (20) give

%rownanie 24
\begin{equation}
m_{I i} \simeq - \frac{\mu^{(L)\,2}_i}{m^{(L)}_s}\;,\; m_{II i} \simeq  
m^{(L)}_s \;,
\end{equation}

\ni where the eigenvalues $\mu^{(L)}_i $ of $\widehat{\mu}^{(L)} = (\mu^{(L)
}_{\alpha \beta})$ are produced with the use of $\widehat{U}_I $:

%rownanie 25
\begin{equation}
\delta_{ij} \mu^{(L)}_i  = \sum_{\alpha \beta} U^*_{I\,\alpha\,i} U_{II\,
\beta\,j} \mu^{(L)}_{\alpha \beta}\;.
\end{equation}

\ni In this case, $\nu_{II \alpha}$ ($\alpha = e\,,\,\mu\,,\,\tau $) are 
(approximate) neutrino heavy mass eigenstates decoupled from $\nu_{I i}$ ($i = 
1,2,3 $).

 Concluding, the lefthanded see--saw, constructed in this note, implies both 
the smallness of active--neutrino masses and decoupling of heavy passive 
neutrinos, similarly to the situation in the case of conventional see--saw. 
However, introducing as heavy passive neutrinos the lefthanded sterile neutr%
inos $\nu_{\alpha L}^{(s)}$ in place of the conventional righthanded neutrinos 
$\nu_{\alpha R}$ ($\alpha = e\,,\,\mu\,,\,\tau $), the new see--saw mechanism, 
when spoiling the chiral left--right pattern of original \SM for neutrinos, 
does it in a way different from the conventional see--saw mechanism. Recall 
that in the lefthanded see--saw the righthanded neutrinos continue to be 
completely absent.

Note finally that in the case of lefthanded see--saw the active and sterile ne%
utrinos, as practically unmixed, cannot oscillate into each other. They could,
if in place of the inequality (10) in Eq. (2) the relation

%rownanie 26
\begin{equation}
\mid m^{(L)} - m^{(L)}_s \mid \ll \mu^{(L)}
\end{equation}

\ni were conjectured, leading to their nearly maximal mixing. This mechanism, 
however, would not imply automatically small neutrino masses, since then $ 
m_{I,II} \simeq (1/2)(m^{(L)}+m^{(L)}_s) \mp \mu^{(L)}$ from Eq. (9).
.

 I am indebted to Stefan Pokorski for several helpful discussions.

~~~~
\vspace{1.0cm}

{\centerline{\bf References}}

\vspace{0.3cm}

{\everypar={\hangindent=0.5truecm}
\parindent=0pt\frenchspacing

{\everypar={\hangindent=0.5truecm}
\parindent=0pt\frenchspacing

~1.~M.~Gell--Mann, P.~Ramond and R.~Slansky, in {\it Supergravity }, ed. 
P.~van Nieuwenhuizen and D.Z.~Freedman, North--Holland, Amsterdam, 1979;
T.~Yanagida, in {\it Proc. of the Workshop on the Unified Theory of the 
Baryon Number in the Universe}, ed. O.~Sawada and A.~Sugamoto, KEK report 
No. 79--18, Tsukuba, Japan, 1979; see also R.~Mohapatra and G.~Senjanovic,
{\it Phys. Rev. Lett.} {\bf 44}, 912 (1980).

\vspace{0.15cm}

~2.~R.D.~Peccei, hep--ph/9906509.

\vspace{0.15cm}

~3.~G.~Altarelli and F.Feruglio, CERN--TH/99--129 + DFPD--99/TH/21, hep--ph/%
9905536.

\vfill\eject

\end{document}